# CMB ANISOTROPIES, LARGE-SCALE STRUCTURE AND THE FUTURE

DOUGLAS SCOTT
Astronomy Department, University of California, Berkeley, CA 94720, U.S.A.
and Department of Geophysics & Astronomy, University of British Columbia, 129-2219 Main Mall, Vancouver, B.C., Canada V6T 1Z1

ABSTRACT  We are now beginning to learn detailed information about cosmological parameters from the shapes of the matter and radiation power spectra, together with their relative normalization. As more high quality data are gathered from galaxy surveys and from microwave anisotropies, the range of allowed models is expected to get incrementally smaller. The amount of information potentially available from a high-resolution satellite experiment should allow a determination of essentially *all* currently discussed cosmological parameters to $\lesssim 10\%$.

## CMB AND LSS POWER SPECTRA

The Hot Big Bang model is now well-established and the current big cosmological puzzles are concerned with where the structures we observe in the Universe came from. What were the 'initial conditions' (i.e. primordial fluctuations), and how were they generated? How did these fluctuations grow, i.e. can we learn about dark matter and other information concerned with the cosmological background? There are two distinct domains from which we can answer such questions: the power spectrum of matter fluctuations; and the power spectrum of radiation fluctuations. A useful way of thinking about these is that for a given input theory (in other words a given list of cosmological parameters) they are *two separate* outputs. Moreover, they are the main outputs, from which we can hope to determine the underlying parameters.

There are many ways of writing these spectra, but the most common is to use the quantitities $P(k)$ and $C_\ell$ (see e.g. White, Scott & Silk 1994). $P(k)$ is the spectrum of squared amplitudes of the fourier expansion of density perturbations $\delta(\mathbf{x})$. For isotropic models only $|\mathbf{k}|$ is important. $C_\ell$ is the spectrum of squared amplitudes of the spherical harmonic expansion ($Y_{\ell m}(\hat{n})$, the analogue of fourier series on a curved sky) of the temperature $T(\hat{n})$. For isotropic models only the $\ell$'s, and not the individual $m$'s, are important. Theoretical calculations of these spectra involve numerical integration of coupled Boltzmann equations for each cosmological fluid (e.g. Bond 1995). These calculations have been developed to the point where they are expected to be accurate at the $\simeq 1\%$ level (see HSSW).

$P(k)$ information comes from the local universe at $z \simeq 0$. Measurements come from clustering of various objects, together with their abundances and their velocity flows and dispersions. $C_\ell$ information comes from radiation fluctuations on the last scattering surface at $z \simeq 1000$, together with some physical processes occurring at lower redshifts. Measurements come from a range of angular sizes from tens of degrees down to arcseconds.

Although there is an overlap in the scales probed by these two approaches, there is rather different physics operating — so there will always be complementary information obtainable. On the largest scales both spectra measure (super-)horizon modes, which means sensitivity to initial conditions, with no subsequent processing by causal effects. The break in $P(k)$ (see Fig. 1) largely measures the redshift of the equality epoch between matter and radiation, which depends on $\Omega_0 h$. The slope of the initial conditions can be determined if there is a large enough lever-arm in $k$. Information can also be obtained on the sort of dark matter (e.g. the fractions of neutrinos and baryons), from the detailed shape around this peak region, and the damping at smaller scales.

The $C_\ell$ spectrum (see Fig. 4) can be used to measure $n$ in the same way, and can also indicate the amount of tensors (gravity waves) if they are a reasonable fraction of the scalars (density perturbations). More importantly though, the physics of sound waves imprints information about gravity and pressure at $z \simeq 1000$ (see e.g. Scott & White 1995), and geometric projection between last scattering and today. This means that the spectrum additionally encodes variations in $\Omega_0$, $\Omega_B$, $\Omega_\Lambda$ and $h$ (as well as $\Omega_\nu$, $\tau_i$, ...).

Furthermore, there is only *one* free normalization for the two power spectra, so that the ratio of normalizations (e.g. the quantity $C_2/\sigma_8$) contains more information. In particular this ratio can determine the tensor fraction $T/S$ and parameters which affect the growth rate between $z = 1000$ and $z = 0$, mainly $\Omega_0$ and $\Omega_\Lambda$.

## THE MATTER POWER SPECTRUM

So where are we with measurements of $P(k)$ and $C_\ell$? Several surveys of galaxies have led to estimates of $P(k)$ (see e.g. Peacock & Dodds 1994, and Fig. 1). Although there is some disparity between the surveys (see Efstathiou 1995), the indications are that the 'shape parameter' $\Gamma$ ($\simeq \Omega_0 h$ if $n = 1$) is significantly less than the value of 0.5 predicted by standard Cold Dark Matter (sCDM: $\Omega_0 = 1$, $h = 0.5$, $n = 1$). A reasonable range for $\Gamma$ is 0.2–0.35. It is important to understand that although an estimate of $\Gamma$ appears to be a determination of the turnover in $P(k)$, in reality current surveys have little sensitivity at such large scales, and so the measurement is really of some slope, i.e. $\Gamma$ is currently equivalent to, say, $\sigma_{50}/\sigma_8$.

We also know that $\sigma_8$ (the rms overdensity in spheres of radius $8\,h^{-1}$Mpc) is significantly less than unity. sCDM gives $\sigma_8 = 1.34$ when normalized to *COBE* (Bunn, Scott & White 1995), which is unquestionably too high. Cluster abundances (e.g. White, Efstathiou & Frenk, 1993) indicate that for total mass $\sigma_8 \simeq (0.5\text{–}0.8)\Omega_0^{-0.6}$. Like the determination of $\Gamma$ this estimate varies somewhat between authors, and the specifics of the technique used. Since clusters probe scales around $8\,h^{-1}$Mpc, the measurement is relatively insensitive to the slope of the spectrum.

There is also information at smaller scales, e.g. from the abundance of high $z$ damped Ly $\alpha$ clouds or quasars, or from the galaxy pairwise relative velocities. Both sets of constraints probe scales $\sim 1 h^{-1} \mathrm{Mpc}$, with the former tending to prefer high values of $\sigma_1$, while the latter prefers lower values. Realistic models presumably lie somewhere in between, where the fit to both is adequate. This region of the spectrum is further complicated because the fluctuations are in the non-linear regime.

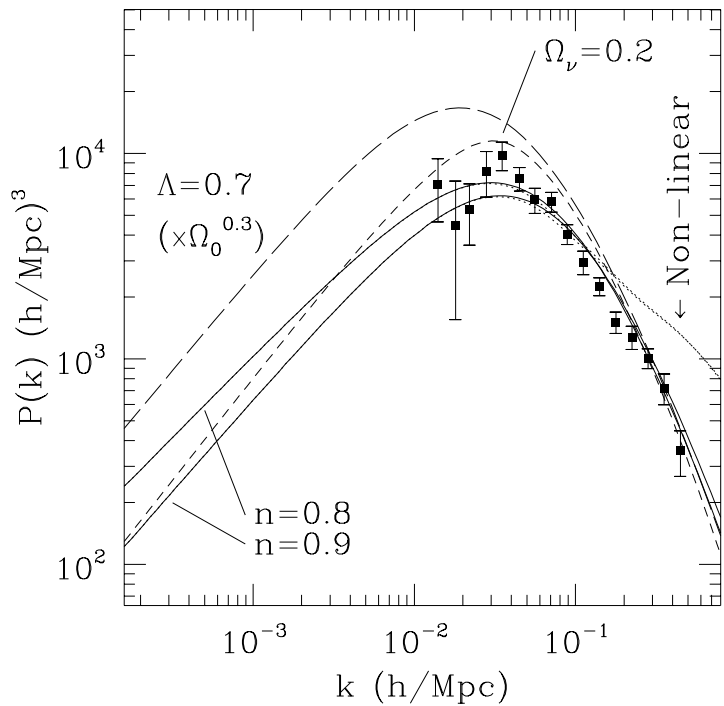

Fig. 1. The matter power spectra for 'tuned' CDM models (solid), along with an MDM model (short dashed) and $\Lambda$CDM model (long dashed) for comparison. The $\Lambda$CDM model has been multiplied by $\Omega_0^{0.3}$ to allow comparison with the data points (from Peacock & Dodds 1994). The non-linear $P(k)$ for the CDM models is shown dotted. [From White et al. 1995].

Fig. 1 shows one attempt to construct a model which adequately fits most of the information we have on $P(k)$ and the CMB, while being accurately normalized to the 2 year *COBE* data (Bennett et al. 1994). This is a 'tuned' CDM model, where we have taken the parameters of sCDM and varied them by $\sim 10\%$ in order to better fit the data (see White et al. 1995). We find that the current 'best-buy' CDM model is either $n = 0.9, h = 0.45$ with tensors, or $n = 0.8, h = 0.45$ without tensors (to quote some round numbers). These models may already be in conflict with estimates of $\Omega_0$ and of $H_0$, although such estimates have always been the subject of dispute.

Note that if the Hubble constant is really $h \gtrsim 0.6$ as more and more recent estimates seem to be suggesting (e.g. Freedman et al. 1994), then all $\Omega_0 = 1$ models are in trouble. If the age crisis continues to strengthen, then we may be forced to consider less appealing models. For example the oldest model for a given $\Omega_0$ has the highest $\Omega_\Lambda$ allowed by other constraints (e.g. quasar lensing) — a specific example of an oldest possible universe is *closed* with $\Omega_0 \simeq 0.4$ and $\Omega_\Lambda \simeq 0.9$ (White & Scott 1995). In any case, more detailed measures of $P(k)$, which are expected to come from the AAT 2dF and the Sloan survey, will presumably shed some light on whether any attractive CDM-like models continue to fit, or whether we need to invoke open models, $\Lambda$ or massive neutrinos (or an entirely different paradigm).

The CMB anisotropies will also be important in constraining such models. Already the possible existence of more power on degree-scales than *COBE* scales (the emergence of the acoustic peak, Scott & White 1994) provides an intriguingly novel limit on the models. If we take sCDM normalized to *COBE*, then something has to be done to reduce $\Gamma$ and $\sigma_8$. The most obvious thing is to tilt the spectrum to $n < 1$ and perhaps to add a tensor contribution at the same time. However, both of these changes will lead to a reduction in the height of the acoustic peak. Hence if the peak-height becomes securely measured at some value, then that provides a tight restriction on how much $n$ can be reduced and $T/S$ increased. Variations in $\Omega_0$, $\Omega_\Lambda$ and $h$ have different effects, so that a combination of the $P(k)$ and $C_\ell$ information provides a complex set of parameter constraints.

## THE RADIATION POWER SPECTRUM

What is the status of CMB anisotropies? Fig. 2 shows one representation of the current experimental situation. It is convenient to express the anisotropies in *amplitude* units, plotted linearly and with symmetric error bars (i.e. Fig. 2 is the square root of the power spectrum of Fig. 4). $Q_{\rm flat}$ is specifically the best-fit amplitude of a flat spectrum (i.e. $\ell(\ell+1)C_\ell =$ const.) for the range of scales probed by each experiment, quoted at the quadrupole scale. This can easily be related to other measures of anisotropy (White & Scott 1994). Note that vertical error bars are $\pm 1\sigma$ for detections and 95% confidence levels for upper limits, while the horizontal bars simply indicate the range of $\ell$ probed by each experiment. Specifics of the experiments are given in Scott & White (1994) and Scott, Silk & White (1995). [This plot includes new points from the Python experiment (Mark Dragovan, private communication), improved numbers for SP91 and SP93 (Josh Gundersen, private communication) and the low MAX point now treated as an upper limit. The open circles at small $\ell$ are the maximum entropy fits to $\ell = 2$–19 for the 2-year *COBE* data, obtained from Ted Bunn (see Bunn 1995)].

If the spectrum *was* flat, then these points would scatter about a straight line. It seems clear already that the data prefer a peak around $\ell \sim 200$ — the main acoustic peak — although clearly a wide range of models are consistent with this collection of data points. However, there are a few things which we have already learned. Firstly, it is worth pointing out that these data are actually fairly consistent with each other and not 'all over the place', as is sometimes suggested. Secondly, we can use the apparent minimum height of the peak to put a constraint on the amount of reionization which is possible. Conservatively assuming a high $\Omega_B$ and no tilt or tensors, the limit to reionization optical depth is $\tau_i < 0.5$. This means that the universe was probably not ionized before $z \simeq 50$. Perhaps not a surprising conclusion, but one that was not previously known.

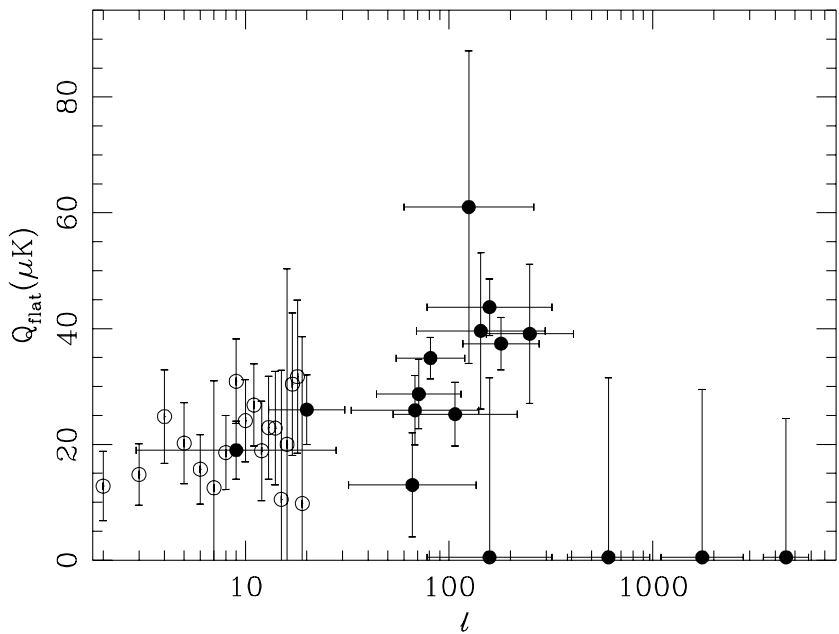

Fig. 2.    The current status of CMB anisotropy experiments. The y-axis is the best-fitting amplitude of a flat spectrum of anisotropies, seen through the window function of each experiment, and quoted at the quadrupole scale. Error bars are $\pm 1\sigma$. The x-axis is multipole, where $\ell \sim \theta^{-1}$. Horizontal bars show the range of scales probed by each experiment. For more details see Scott & White (1994) and Scott, Silk & White (1995). In addition there are some up-dated points for the ACME and Python experiments, and the open circles at small $\ell$ are estimators for *COBE* multipoles 2–19 from Bunn (1995).

We can also say that at the moment the best-fitting model is something roughly like CDM. This means that reasonable values of $\Omega_B$ provide a good fit to the peak height. It also means that very low $\Omega_0$ models are not a good fit to today's data (the peak would be too far to the right). To be quantitative, it is hard to find a curve which goes through the data unless $\Omega_0 \gtrsim 0.3$ (Scott, Silk & White 1995). This is essentially the same limit on $\Omega_0$ which has been obtained from velocity data (e.g. Dekel 1994) and from the *COBE* data alone (Yamamoto & Bunn 1995).

Similarly, isocurvature models are not a very good fit (and don't fit the *COBE* data very well either, see Hu, Bunn & Sugiyama 1995), and the best fitting isocurvature models have to be tuned to have $C_\ell$'s which look as much like CDM as possible! Predictions for defect models (cosmic strings or textures) are currently in a state of flux. It seems clear that there will be some sort of enhancement on degree-scales, but it is unclear whether there will be a series of acoustic peaks or just one smoothed out peak (Albrecht et al. 1995, Crittenden & Turok 1995, Durrer, Gangui & Sakellariadou 1995). Current understanding suggests that the answer will depend on the specific sort of defect, meaning that even if defects are the right answer the $C_\ell$'s will probably be the best way to get more information.

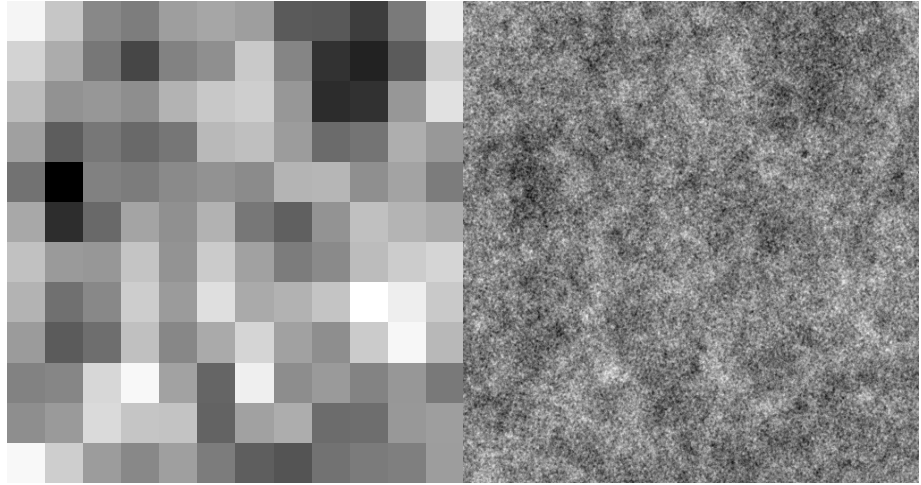

Fig. 3. This is a 30° × 30° patch of the microwave sky for a standard Cold Dark Matter spectrum. The map on the left is smoothed with the *COBE* resolution and binned into *COBE*'s 2°.6 × 2°.6 pixels. The map on the right shows the same patch of sky viewed by a satellite with 10′ resolution and the same degree of oversampling. Note that there are actually pixels plotted here!

## THE FUTURE

As far as the CMB is concerned the subject is likely to be driven by new data. Essentially all the experiments represented on Fig. 2 have more data which is right now being analysed or at least taken. And most of the groups are developing improved versions for the coming 2 or 3 years. On about that same timescale there is the promise of a new generation of longer duration balloon flights, plus large dedicated interferometers. These should lead to some determination of the shape of the power spectrum, and perhaps the height and position of the main acoustic peak. We will certainly have strong constraints on open models (or detections!), since they tend to have lots of small-scale power (e.g. Hu & White 1995). But the really exciting prospect for the future lies in the possibility of a new satellite experiment with sub-degree resolution. Already today's detectors are 100 times more sensitive than those which flew on the

*COBE* mission (which means the same signal-to-noise in one ten-thousandth of the time). There is some debate about whether HEMT's or bolometers are the detectors of choice. And there is a question about whether any of the U.S. efforts (*FIRE, MAP, PSI*) or the European effort (*COBRAS/SAMBA*) will be successful in receiving funding. But whatever eventually flies, the potential for learning fundamental things about the Universe is awe-inspiring. Fig. 3 illustrates the difference between the *COBE* experiment and a satellite experiment with 10′ beam-size. With such resolution it becomes possible to resolve the characteristic scale of the anisotropies ($\sim 0°.5$ for $\Omega = 1$), enabling us to decode the cosmological information with extraordinary accuracy.

Fig. 4 illustrates another way of seeing how much we gain from a new satellite. The theoretical models shown are all ones which are currently under discussion as more or less acceptable ways of fitting most of today's data, with the sCDM model shown for reference. The specific models are sCDM, an open CDM model and a best-fitting $\Lambda$ model (see Scott, Silk & White 1995), together with a preliminary calculation for cosmic strings (from Albrecht et al. 1995). There will be so much good information from such a wide range of $\ell$ that the satellite will be able to distinguish between models which are *much* more similar than the models shown here.

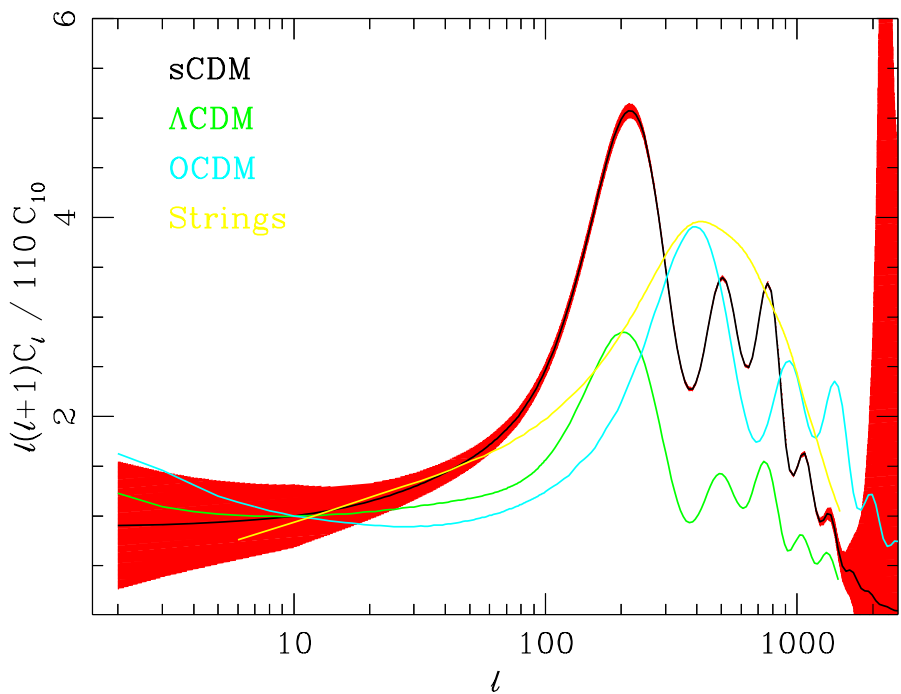

Fig. 4.— The lines are 4 representative models which fit most of the current cosmological data. The band shows the expected error on the $C_\ell$'s for an experiment with 10′ beam-size and 10 $\mu$K noise, and with 10% averaging in $\ell$ for the purposes of this plot. It is clear that these theories could easily be distinguished.

It is possible to estimate how well specific cosmological parameters can be determined by such a satellite. This task is complicated by the foreground contamination which will have to be reduced by multi-frequency measurements. It is possible to show (following Brandt et al. 1994) that foregrounds with no angular correlations will lead to increased noise in only an rms sense, and that any extra angular information only makes them easier to remove. Uncertainties in fitting parameters can then be estimated by assuming a beam-size, noise (*including* an estimate of the effect of having to remove foregrounds), and fraction of sky from which the foregrounds are removable.

For a specific set of experimental parameters, the error on individual $C_\ell$'s is a combination of the 'cosmic variance' (because our sky is only one sample of a hypothetical ensemble) and the contribution from the noise per pixel. This expression can be written as

$$\left(\frac{\delta C_\ell}{C_\ell}\right) = \sqrt{\frac{2}{2\ell+1}} \left(1 + \frac{4\pi}{N} \frac{\sigma_\ell^2}{C_\ell W_\ell}\right),$$

where $N$ is the total number of pixels, $\sigma_\ell$ is the noise in multipole $\ell$ (which would be a constant if the sky were uniformly sampled) and $W_\ell \simeq \exp\left(-\ell^2 \theta_{\rm beam}^2\right)$ is the window function. To generalise this to finite sky you simply multiply the whole expression by $\sqrt{4\pi/A}$ to increase the 'cosmic variance' to the 'sample variance' (Scott, Srednicki & White 1994), and then replace the $4\pi$ inside the brackets with $A$ to account for the increased pixel noise. Here $A$ is the solid angle covered by the experiment, or the foreground-free part that is being analysed. This formula will be *approximately* correct for reasonably large fractions of the sky, with some caveats: it will clearly fail for $\ell \lesssim \sqrt{1/A}$ or for $\ell \gtrsim \sqrt{N}$; there will be some correlations induced between the $C_\ell$'s which ought to be taken into account; and the specific effect on individual $C_\ell$'s (particularly at low $\ell$) will depend on the particular shape of the area covered.

Qualitatively this formula was used to derive the error band of Fig. 4, where the input theory was sCDM. Further simulations showed that it is possible to estimate $\Omega_0$, $\Omega_B$ and $n$ to a few percent, depending little on their input values, provided the beam and noise are not too dissimilar from what we assumed in Fig. 4. The fraction of gravity waves can be measured relatively well providing they are $\gtrsim 20\%$ of the scalars. There are only a few other parameters in the average cosmologists portfolio: $\Lambda$, $h$ and $\tau_i$. For most of their possible ranges these can be measured to unprecedented accuracy. Together with the overall normalization, this makes 7 free parameters to be fitted. In addition the CMB power spectrum obtained from a future satellite mission should be so well determined that we are sensitive to some effects which have normally been neglected up until now (see HSSW for a compendium of many 1% effects). Theorists may have to be more industrious and imaginative to keep up with this sort of data!

ACKNOWLEDGEMENTS

Essentially all of the work discussed here has been done in collaboration with Martin White, and is reported in a series of papers written with him during my time in Berkeley. I would also like to thank the other members of the now defunct Berkeley CMB group for countless invaluable discussions: Joe Silk, Ted Bunn, Wayne Hu and Naoshi Sugiyama.

REFERENCES

Albrecht, A., Coulson, D., Ferreira, P. & Magueijo, J. 1995, Imperial preprint, astro-ph/9505030.
Bennett, C. L., et al., 1994, *Ap. J.*, **436**, 423.
Bond, J. R., 1995, in *Les Houches Lectures 1993*, ed. R. Schaeffer, Elsevier Science Publishers, Netherlands, in press.
Brandt, W. N., Lawrence, C. R., Readhead, A. C. S., Pakianathan, J. N. & Fiola, T. M. 1994, *Ap. J.*, **424**, 1.
Bunn, E. F. 1995, *Ph. D. thesis*, University of California, Berkeley.
Bunn, E. F., Scott, D. & White, M. 1995, *Ap. J.*, **441**, L9.
Bunn, E. F. & Sugiyama, N. 1995, *Ap. J.*, **446**, 49.
Crittenden, R. G. & Turok, N. G. 1995, Princeton preprint, astro-ph/9505120.
Dekel, A., 1994, *Ann. Rev. Astron. Astrophys.*, **32**, 371.
Durrer, R., Gangui, A. & Sakellariadou, M. 1995, Zürich preprint, astro-ph/9507035.
Efstathiou, G. 1995, in *Les Houches Lectures 1993*, ed. R. Schaeffer, Elsevier Science Publishers, Netherlands, in press.
Freedman, W. L., et al., 1994, Nature, **371**, 757.
Hu, W., Bunn, E. F. & Sugiyama, N. 1995, *Ap. J.*, **447**, L59.
Hu, W., Scott, D., Sugiyama, N. & White, M. 1995, *Phys. Rev. D*, in press, astro-ph/9505043 (HSSW).
Hu, W. & White, M. 1995, *Astr. Ap.*, in press, astro-ph/9507060.
Peacock, J. A. & Dodds, S. J. 1994, *M.N.R.A.S.*, **267**, 1020.
Scott, D., Srednicki, M. & White, M. 1994, *Ap. J.*, **421**, L5.
Scott, D., Silk, J. & White, M. 1995, *Science*, **268**, 829.
Scott, D. & White, M. 1994, in *CMB Anisotropies Two Years After COBE*, ed. Lawrence Krauss, World Scientific, Singapore, p. 214.
Scott, D. & White, M. 1995, *Gen. Rel. & Grav.*, in press, astro-ph/9505102.
White. M. & Scott, D. 1994, in *CMB Anisotropies Two Years After COBE* ed. Lawrence Krauss, World Scientific, Singapore, p. 254.
White, M. & Scott, D. 1995, "*Why not consider closed universes?*", *Ap. J.*, submitted, astro-ph/9508157.
White, M., Scott, D. & Silk, J. 1994, *Ann. Rev. Astron. Astrophys.*, **32**, 319.
White, M., Scott, D., Silk, J. & Davis, M. 1995, "*Cold Dark Matter Resuscitated?*", *M.N.R.A.S.*, in press, astro-ph/9508009.
White, S. D. M., Efstathiou, G. & Frenk, C. S., 1993, *M.N.R.A.S.*, , **262**, 1023.
Yamamoto, K. & Bunn, E. F. 1995, Berkeley preprint, astro-ph/9508090.